\documentclass{moriond}

\def\Journal#1#2#3#4{{#1} {\bf #2}, #3 (#4)}

\def\NPB{{\em Nucl. Phys.} B}
\def\PLB{{\em Phys. Lett.}  B}
\def\PRL{\em Phys. Rev. Lett.}
\def\PRD{{\em Phys. Rev.} D}

\def\EPJC{{\em Eur. Phys. J.} C}

\def\be{\begin{equation}}
\def\ee{\end{equation}}
\def\bea{\begin{eqnarray}}
\def\eea{\end{eqnarray}}

\newcommand{\Photo}{}

\begin{document}
\vspace*{4cm}
\title{Revisiting strong-coupling determinations from $e^+ e^-$ event shapes}

\author{Guido Bell\,\textsuperscript{\textit{a}}, Christopher Lee\,\textsuperscript{\textit{b}}, Yiannis Makris\,\textsuperscript{\textit{b,c}}, Jim Talbert\,\textsuperscript{\textit{d,b}}, Bin Yan\,\textsuperscript{\textit{b,e}}}
\address{\textsuperscript{\textit{a}}\,Theoretische Physik 1, Center for Particle Physics Siegen, Universit\"at Siegen, 57068 Siegen, Germany\\
\textsuperscript{\textit{b}}\,Theoretical Division, Los Alamos National Laboratory, Los Alamos, NM  87545, USA\\
\textsuperscript{\textit{c}}\,INFN Sezione di Pavia, 27100 Pavia, Italy\\
\textsuperscript{\textit{d}}\,Institute for Defense Analyses, Alexandria, VA 22305, USA\\
\textsuperscript{\textit{e}}\,Institute of High Energy Physics, Chinese Academy of Sciences, Beijing 100049, China
}

\maketitle\abstracts{
We reassess the theoretical uncertainties of strong-coupling determinations from a global fit to the $e^+ e^-$ thrust data, focusing in particular on the scheme dependence associated with the applied renormalon-cancellation prescription and on the choice of scale parameters that are used to estimate higher-order perturbative corrections in all sectors of the calculation.}

\section{Introduction}

Electron-positron event shapes are amongst the oldest and most established observables that are used to determine the strong-coupling constant $\alpha_s$. The most precise extractions from a global fit to the thrust~\cite{Abbate:2010xh} and C-parameter~\cite{Hoang:2015hka} distributions revealed, however, a surprising discrepancy with the world average~\cite{PDG}. This has motivated studies of additional sources of non-perturbative corrections associated with three-jet configurations~\cite{Luisoni:2020efy,Caola:2022vea,Nason:2023asn}. While these exploratory studies are based on specific models, we report on a recent work~\cite{Bell:2023dqs} here, in which we scrutinised elements of the \emph{dijet predictions} that are theoretically very well established. 

\section{Perturbative distribution}

The differential thrust distribution was computed to $\mathcal{O}(\alpha_s^3)$ in the strong-coupling expansion more than 15 years ago~\cite{Gehrmann-DeRidder:2007vsv,Weinzierl:2009ms}. For small values of $\tau=1-T$, the fixed-order prediction fails to provide an adequate description of the distribution due to Sudakov-type corrections that are induced by soft and collinear emissions. The starting point for resumming these corrections to all orders is a dijet factorisation theorem
\begin{equation}
\label{eq:factorisation}
\frac{1}{\sigma_0} \frac{d\sigma}{d\tau} \simeq
H(Q,\mu) \int \! d\tau_n \,d\tau_{\bar n} \,d\tau_s \;
J(\sqrt{\tau_n}Q,\mu)\,J(\sqrt{\tau_{\bar n}}Q,\mu)\, S(\tau_s Q,\mu)\, 
\delta(\tau-\tau_n-\tau_{\bar n}-\tau_s)\,,
\end{equation}
which organises the contributions from hard ($H$), collinear jet ($J$), and soft ($S$) emissions. The factorisation theorem is valid at leading power in $\tau\ll 1$. As all the ingredients are known to sufficiently high order, the Sudakov corrections can be resummed to fourth order (N$^3$LL$^\prime$) in renormalisation-group-improved perturbation theory using effective-field-theory methods. More precisely, there is a single coefficient at this order -- the three-loop soft constant $c_{\tilde{S}}^3$ -- that is currently unknown, and for which we use two different estimates~\cite{Abbate:2010xh,Bruser:2018rad} to study its impact on the $\alpha_s$ extraction.

\begin{figure}
\vspace{-.5em}
    \hspace{5mm}
 \begin{minipage}{0.4\linewidth}
		\centerline{\includegraphics[width=1.05\linewidth]{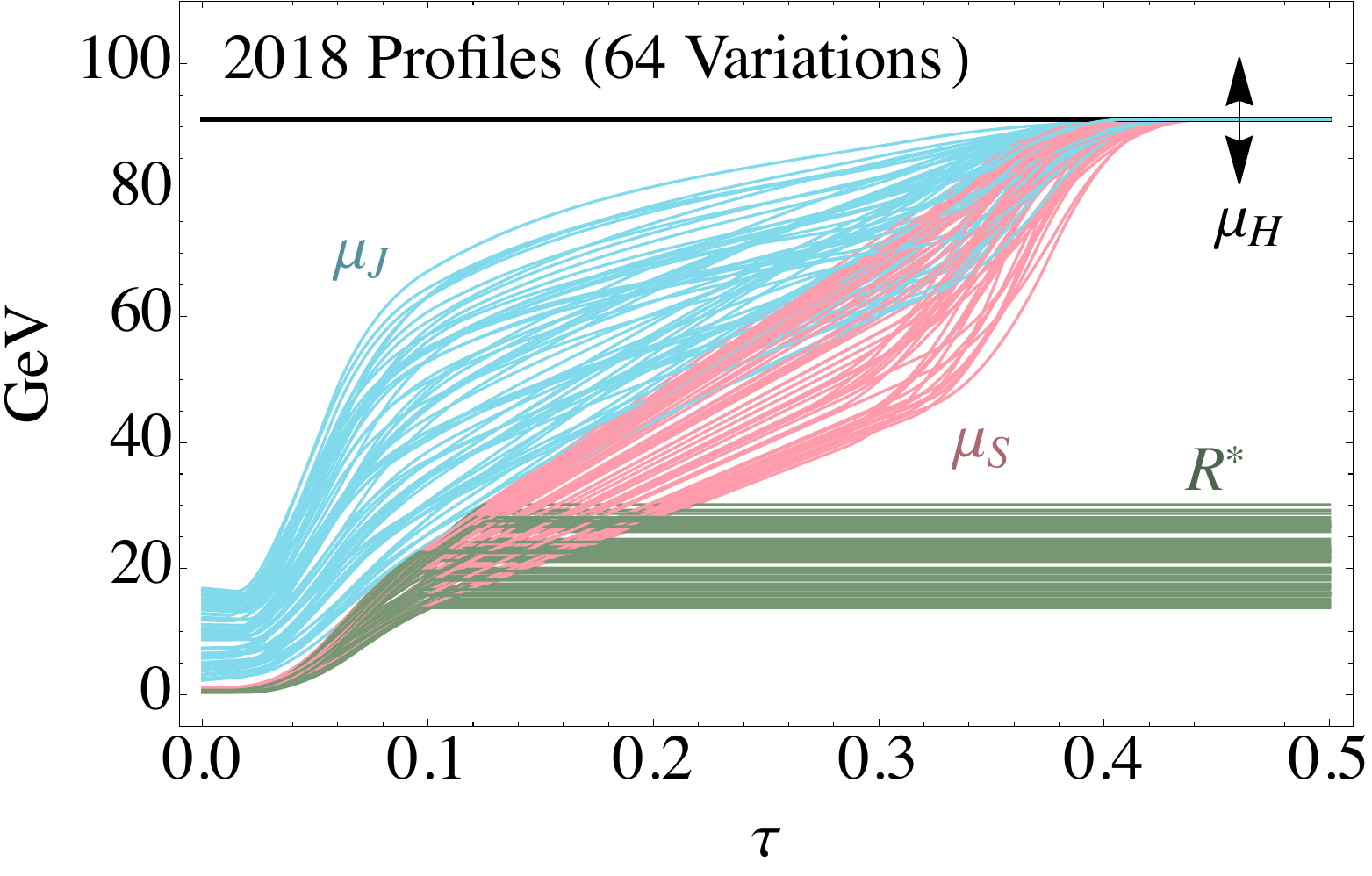}}
	\end{minipage}
	\hspace{10mm}
	\begin{minipage}{0.4\linewidth}
		\centerline{\includegraphics[width=1.05\linewidth]{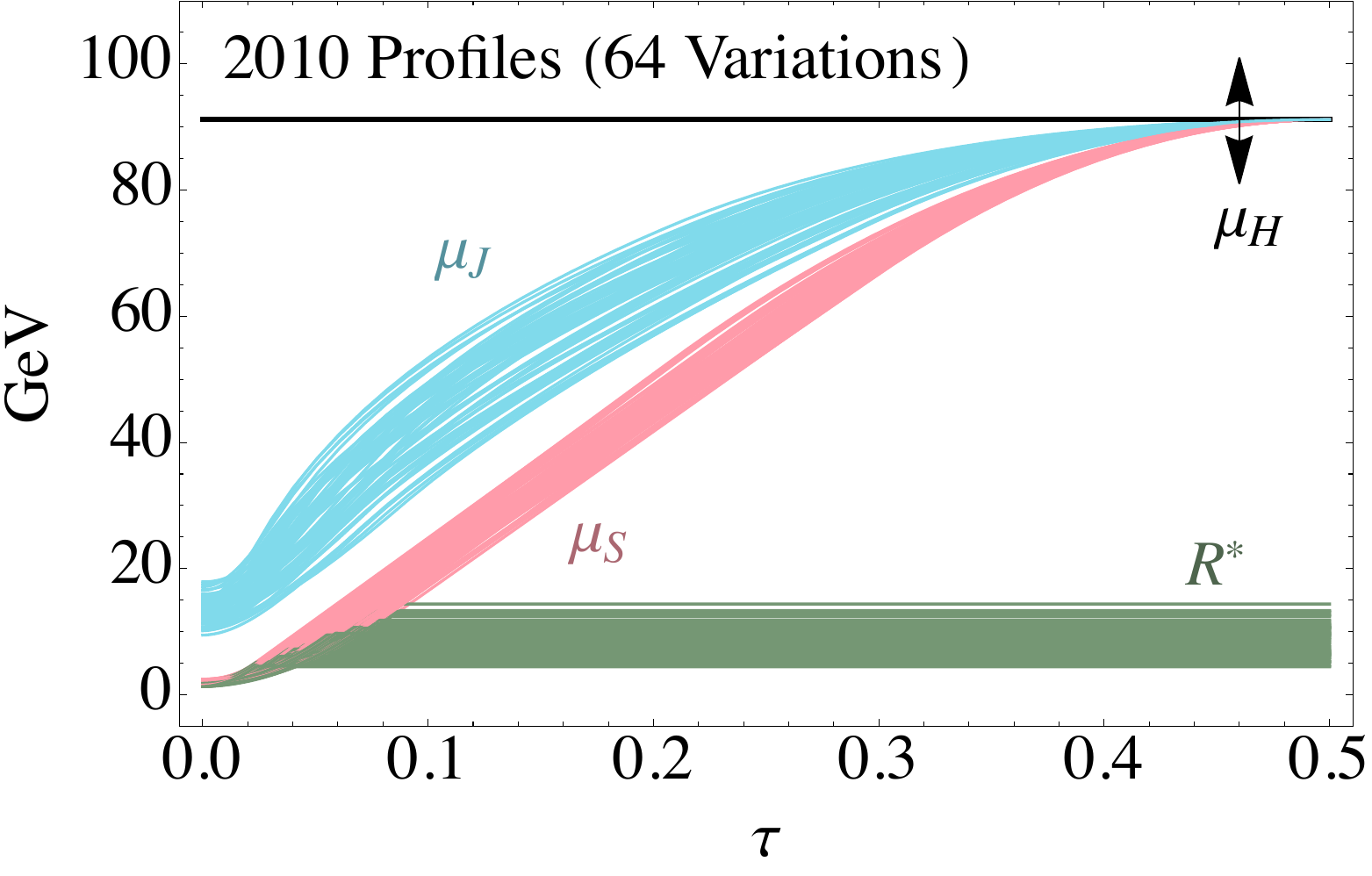}}
	\end{minipage}
 \hspace{5mm}\\[.5em]
 \hspace*{5mm}
	\begin{minipage}{0.4\linewidth}
		\centerline{\includegraphics[width=1.05\linewidth]{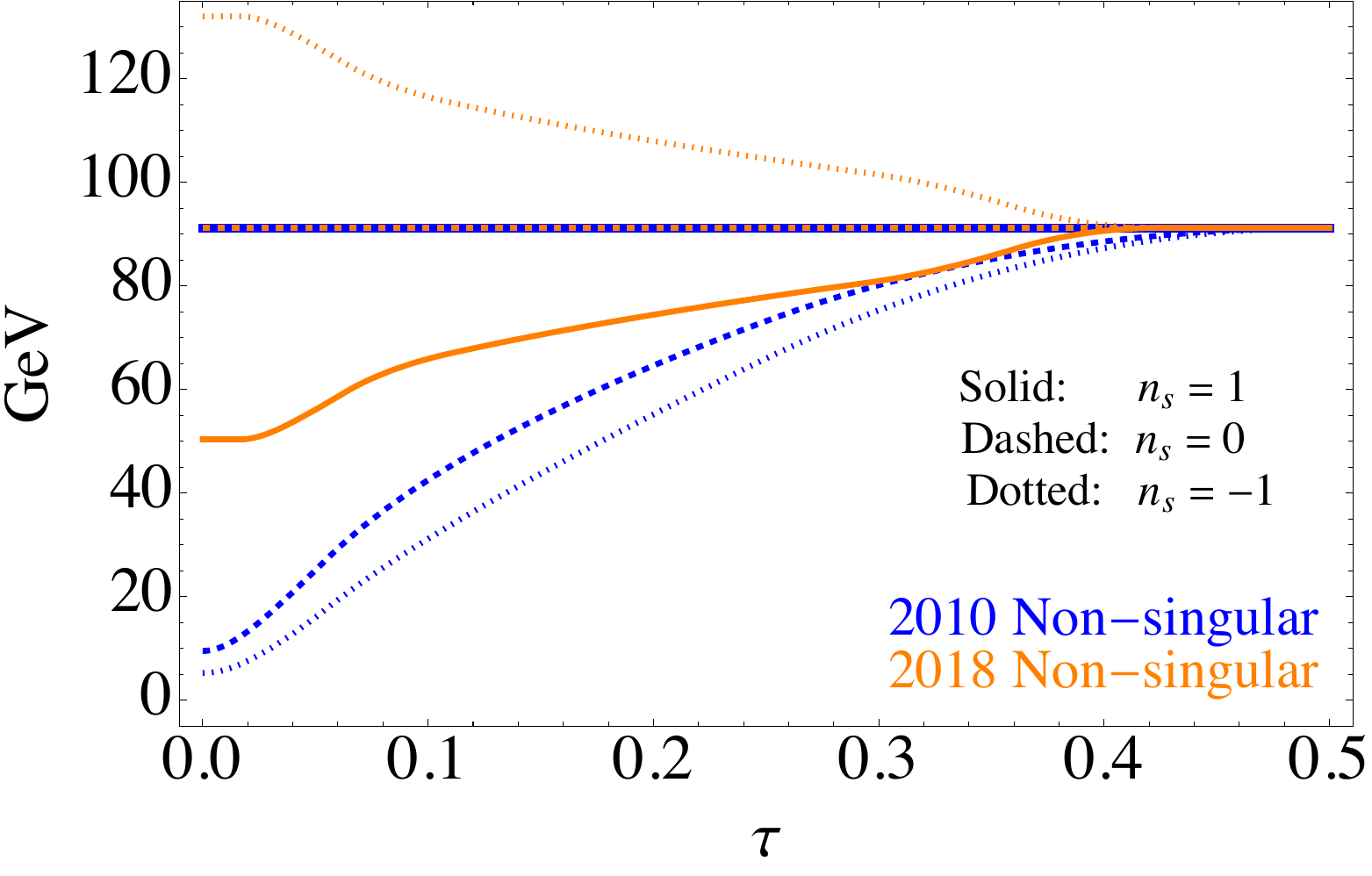}}
	\end{minipage}
	\hspace{11.5mm}
	\begin{minipage}{0.4\linewidth}
		\centerline{\includegraphics[width=1.00\linewidth]{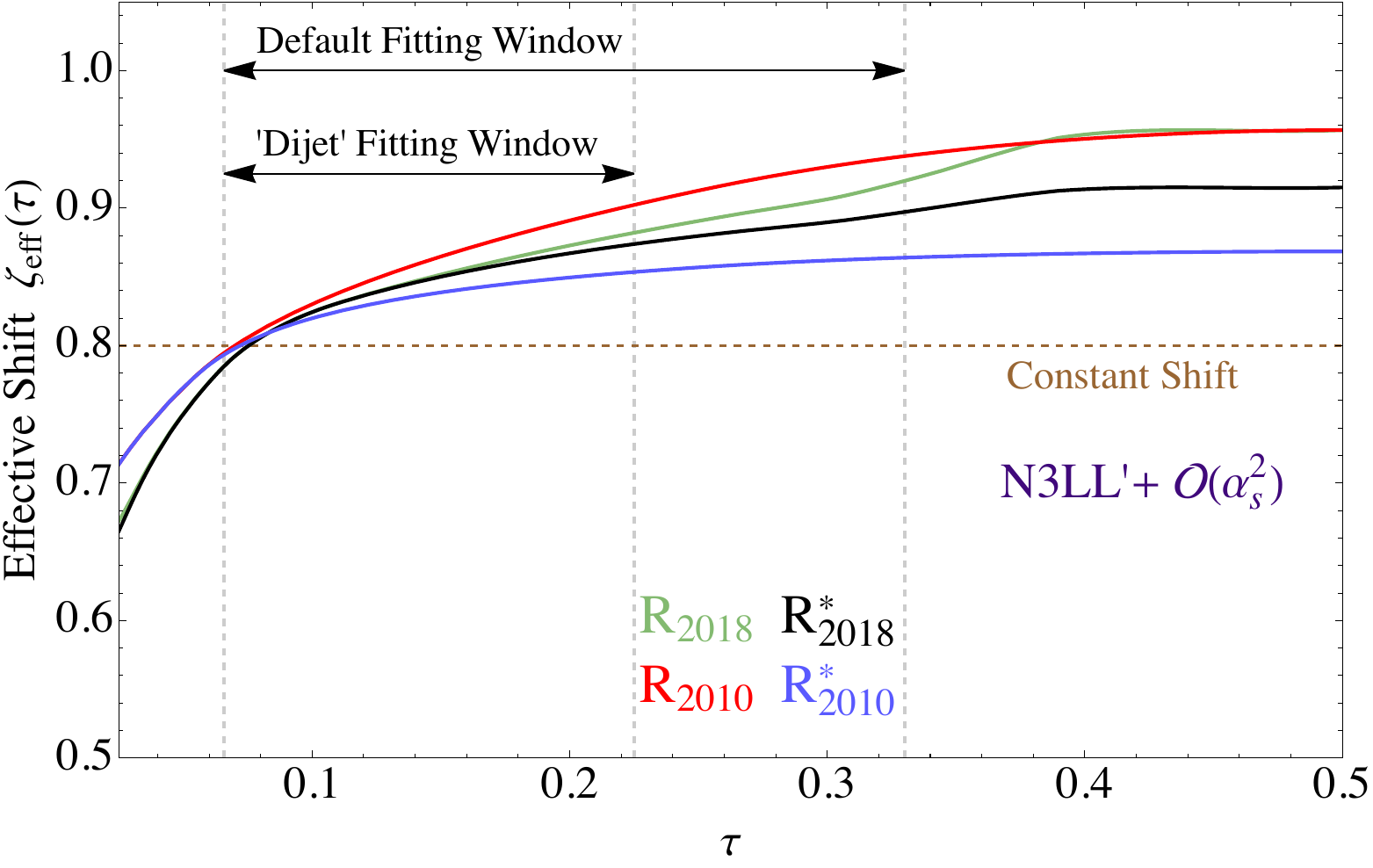}}
	\end{minipage}
    \hspace{5mm}
    \vspace{-.5em}
	\caption[]{First three panels: The 2018 and 2010 profile functions and their variations. Last panel: The effective shift (\ref{eq:effectiveshift}) for different renormalon-cancellation schemes and perturbative scale choices.}
	\label{fig:profiles+shift}
\end{figure}

In order to obtain an accurate description of the entire distribution, the resummed expression is matched onto the fixed-order prediction. As we observed some instabilities in the \texttt{EERAD3}  fixed-order code~\cite{Gehrmann-DeRidder:2014hxk}, we decided to perform the matching at N$^3$LL$^\prime$ + $\mathcal{O}(\alpha_s^2)$ accuracy, but we stress that none of our conclusions are affected by this restriction.\footnote{$\mathcal{O}(\alpha_s^3)$ predictions from \texttt{ColorfulNNLO} are available~\cite{DelDuca:2016ily} but not yet included in our analysis.} In total, the procedure introduces four scales -- three resummation scales $\mu_{H,J,S}$ and a fixed-order (``non-singular") scale $\mu_{ns}$ -- which we vary in order to estimate unknown perturbative corrections in all sectors of the calculation. These scales are dynamical, and the \emph{profiles} we use for them are showcased in the first three panels of Fig.~\ref{fig:profiles+shift}. In particular, the set of ``2018" scales emerged from a simultaneous analysis of seven angularities in one of our previous works~\cite{Bell:2018gce} in an effort to globally optimise the convergence of the uncertainty bands. Comparing them against the ``2010" profiles that were used in past $\alpha_s$ determinations~\cite{Abbate:2010xh}, it is obvious that our scale choices are significantly more conservative.

\section{Non-perturbative effects}

The dijet factorisation theorem (\ref{eq:factorisation}) provides a starting point for implementing non-perturbative corrections in a model-independent way. In particular, the dominant linear corrections are associated with the soft function $S$, and their net effect results in a shift of the perturbative distribution~\cite{Dokshitzer:1995zt,Korchemsky:1999kt}. This shift is driven by a universal non-perturbative parameter~\cite{Lee:2006nr}, a vacuum matrix element of transverse energy flow between soft Wilson lines,  
$\Omega_1 = \frac{1}{N_c}  \mathrm{Tr} \, \big\langle \Omega \big| S^{\dagger}_{\bar{n}} S_{n} \,\mathcal{E}_T\, S^{\dagger}_{n} S_{\bar{n}} \big| \Omega    \big\rangle$, which we extract in conjunction with $\alpha_s$ from the experimental data.

More specifically, the non-perturbative shift is implemented via a convolution of the perturbative soft function $S_{PT}$ with a gapped shape function $f_{\rm mod}$\,,
\begin{equation}
	\label{eq:softfunction}
	S(k,\mu_S) = \int dk' \,S_{PT}(k-k',\mu_S)\,f_{\rm mod}(k'-2\overline\Delta)\,,
\end{equation}
where $\overline\Delta$ is a gap parameter that models the minimum momentum of a hadronic final state. In the relevant tail region of the distribution, the shift is then driven by the first moment of $f_{\rm mod}$. Moreover, both the perturbative soft function $S_{PT}$ and the gap parameter $\overline\Delta$ are known to suffer from renormalon ambiguities in the $\overline{\rm MS}$ scheme~\cite{Hoang:2007vb}. One therefore redefines the gap parameter
\begin{equation}
	\overline{\Delta} = \Delta(\mu_\delta,\mu_R) + \delta(\mu_{\delta},\mu_R),
\end{equation}
in a way that $\Delta$ becomes free of the leading soft renormalon, whereas the subtraction term $\delta$ is chosen to cancel the corresponding ambiguity in $S_{PT}$. We remark that this procedure is not unique, and that it introduces two further scales $\mu_{\delta}$ and $\mu_R$ associated with choices of the renormalization scale $\mu_\delta$ in $S_{PT}$ and momentum scale $\mu_R$ at which the subtraction is performed.

Without going into further details here, we mention that a generalised class of renormalon schemes was put forward in the past~\cite{Bachu:2020nqn}, but the related scheme dependence of the $\alpha_s$ determinations has not been addressed yet. Specifically, the so-called R-gap scheme~\cite{Hoang:2008fs} was used in previous $\alpha_s$ extractions~\cite{Abbate:2010xh,Hoang:2015hka}. In order to quantify the impact of the renormalon subtraction, we show in the last panel of Fig.~\ref{fig:profiles+shift} the \emph{effective shift} of the perturbative distribution, 
\begin{equation}
	\label{eq:effectiveshift}
	\zeta_{\rm eff}(\tau)\equiv \int dk \, k \left[ e^{-2\delta(\mu_\delta,\mu_R)\frac{d}{d k}} f_{\rm mod}\big(k-2 \Delta(\mu_\delta,\mu_R)\big) \right]\,,
\end{equation}
i.e.~the first moment of the gapped shape function \emph{after} renormalon subtraction. Interestingly, one observes that for both sets of scale choices, the effective shift in the R-gap scheme (red and green bands) \emph{grows} for larger values of $\tau$, even into the region where the applied dijet formalism loses its validity. This motivated us to introduce a closely related R$^\star$ scheme, which is similar to the R-gap scheme, except that it treats the scales $\mu_{\delta}$ and $\mu_R$ differently. Most importantly, the subtraction scale $\mu_R=R^*$, which in the original R-gap scheme tracks the soft scale $\mu_S$ in the relevant tail region, is \emph{frozen} in the novel R$^\star$ scheme at the onset of the resummation region, as illustrated by the green bands in the first two panels of Fig.~\ref{fig:profiles+shift}. By doing so, one observes in the last panel of this figure that the effective shift gets mitigated for both scale choices (black and blue bands) by construction. We stress that none of these schemes is preferred \emph{a priori} on theoretical grounds, but varying between them allows us to study if the $\alpha_s$ fits are stable under a $5-10\%$ variation of the leading non-perturbative correction.

\section{Results and discussion}

\begin{figure}
\vspace{-.5em}
    \hspace{2mm}
	\begin{minipage}{0.45\linewidth}
		\centerline{\includegraphics[width=1.00\linewidth]{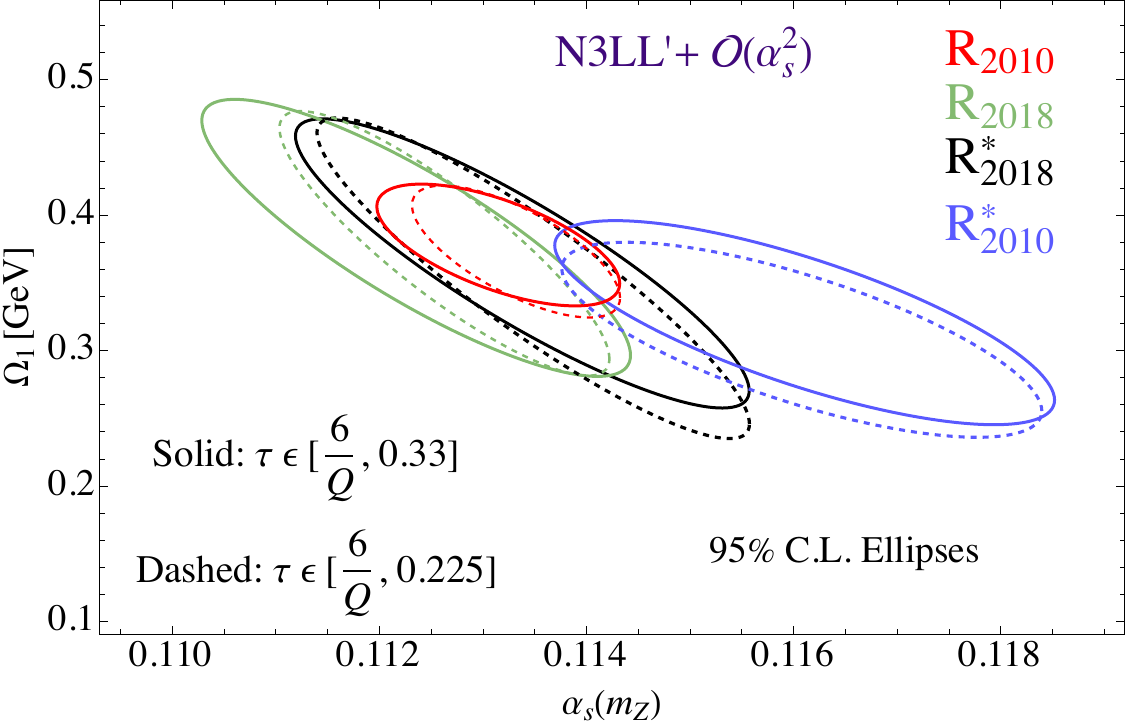}}
	\end{minipage}
    \hspace{5mm}
	\begin{minipage}{0.45\linewidth}
		\centerline{\includegraphics[width=1.00\linewidth]{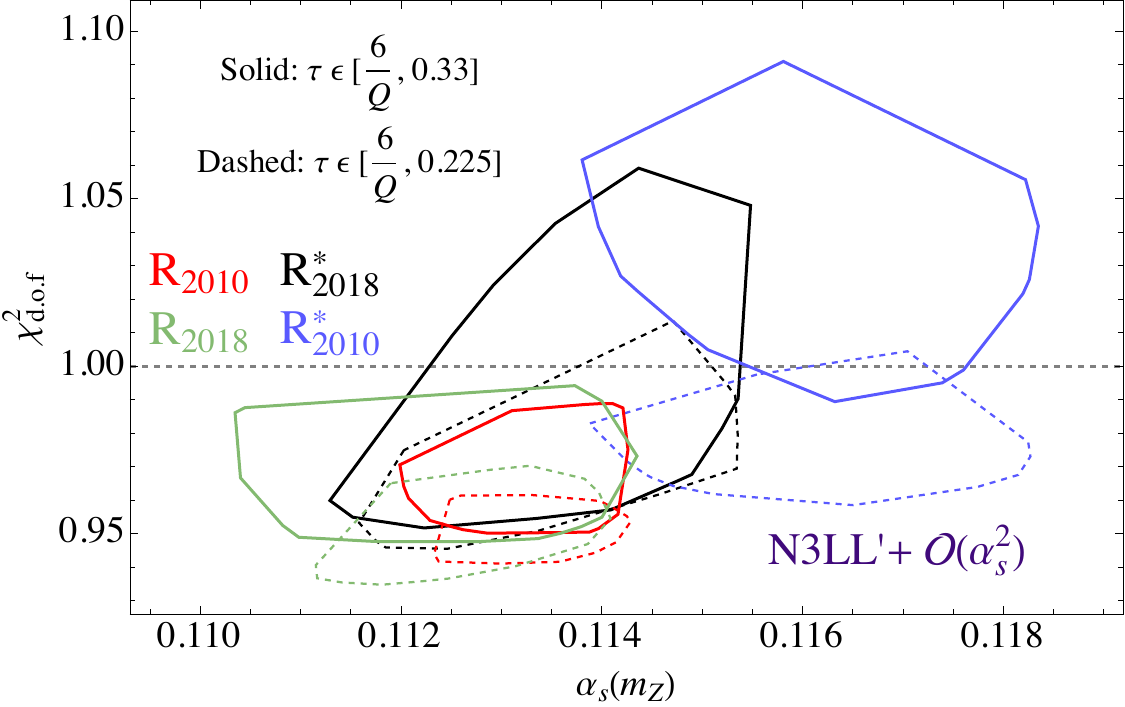}}
	\end{minipage}
     \hspace{2mm}
     \vspace{-.5em}
	\caption[]{Final fit results in the $\alpha_s - \Omega_1$ plane (left) and $\alpha_s - \chi^2_{\rm{dof}}$ plane (right) for four different renormalon-cancellation and scale-variation schemes.}
	\label{fig:results}
\end{figure}

Using the global thrust data, we performed a $\chi^2$-fit on the level of binned distributions to simultaneously extract $\alpha_s\equiv \alpha_s (m_Z)$ and the dominant non-perturbative parameter $\Omega_1\equiv\Omega_1(\mu_\delta,\mu_R)$ at $\mu_\delta = \mu_R = 1.5$~GeV. Particularly, we studied two fit windows $6/Q \le \tau \le 0.33$ (default) and $6/Q \le \tau \le 0.225$ (dijet) based on 52 datasets with center-of-mass energies $Q \in \lbrace 35, 207 \rbrace$ GeV. 

The results are shown in the left panel of Fig.~\ref{fig:results} for two different renormalon-cancellation schemes (R,R$^\star$) and two different scale choices (2018,2010), where the solid (dashed) contours refer to the default (dijet) fit window. Focusing first on the solid contours, we see that our result in the R$_{2010}$ scheme (in red), which mimics the setup that was used in past $\alpha_s$ determinations and confirms their numerical result~\cite{Abbate:2010xh}, yields a low value of $\alpha_s$ with a small uncertainty that is in tension with the world average $\alpha_s(m_Z) = 0.1179 \pm 0.0009$. The uncertainties in the remaining three schemes are, on the other hand, noticeably larger than the red ellipse alone \footnote{Note that $\Omega_1$ is scheme-dependent, and one should therefore not misinterpret the vertical axis in this plot.}. Moreover, one notices that the scheme dependence is more pronounced for 2010 scales than for 2018, which is related to the lower threshold that is used for freezing the subtraction scale $R^*$ (see Fig.~\ref{fig:profiles+shift}). In the right panel of Fig.~\ref{fig:results} we display the corresponding  $\alpha_s - \chi^2_{\rm{dof}}$ plane, which shows that all schemes provide a good fit to the data, with the R$^*_{2010}$ scheme (in blue), which yields the largest $\alpha_s$ value, being slightly less preferred than the others. 

Our analysis reveals that $\alpha_s$ extractions from a fit to the global thrust data can be sensitive to renormalon-cancellation scheme and perturbative scale choices, which we interpret as an indication of additional systematic theory uncertainties. Turning next to the dashed contours in Fig.~\ref{fig:results}, we see in the left panel that the overall pattern in the $\alpha_s - \Omega_1$ plane changes ever so slightly if one fits to data in the more purely dijet-like region, the ellipses moving just slightly but not significantly tighter in $\alpha_s$. More importantly, the right panel in this figure indicates that the fit quality improves noticeably among all considered schemes if the fits are performed in a narrower window.  We therefore advocate that future fits should focus more on dijet-like events in the absence of a model-independent understanding of non-perturbative corrections associated with three-jet configurations (which are most relevant in the far-tail region of the distribution).

We refer to the full publication~\cite{Bell:2023dqs} for further details of our analysis, in particular for a one-to-one comparison of our setup with the one that was used for past $\alpha_s$ determinations~\cite{Abbate:2010xh,Hoang:2015hka}, and for a study of the impact of the currently unknown three-loop soft constant $c_{\tilde{S}}^3$.

\section*{Acknowledgments}

This work was supported by the U.S.~Department of Energy Office of Nuclear Physics, the LDRD and Institutional Computing Programs at LANL, the German Research Foundation under grant 396021762 - TRR 257, and the European  Union’s Horizon 2020 research and innovation programme under the Marie Sk\l{}odowska-Curie grant agreement No. 101022203. Preprint numbers: LA-UR-24-25558, SI-HEP-2024-13, P3H-24-035.

\end{document}